\documentclass[reprint,amsmath,amssymb,aps,prl,superscriptaddress,showkeys]{revtex4-2}

\usepackage{hyperref}
\usepackage{lineno}
\usepackage{graphicx}
\usepackage{natbib}

\hypersetup{
	pdftitle={Giant enhancement of tunnel ionization competes with disorder-driven decoherence in amorphized silicon},
	pdfauthor={D. N. Purschke et al.},
	pdfkeywords={attosecond, disorder, decoherence, tunnel ionization, lightwave electronics, petahertz electronics},
}

\newcommand{\um}{$\mathrm{\mu m}$}

\begin{document}

\title{Giant enhancement of tunnel ionization competes with disorder-driven decoherence in amorphized silicon}


\author{D. N. Purschke}
\email{dpur@lle.rochester.edu}
\affiliation{Joint Attosecond Science Laboratory, National Research Council and University of Ottawa, Ottawa, Ontario K1N 5A2, Canada}
\affiliation{Laboratory for Laser Energetics, University of Rochester, Rochester, New York 14623, United States}

\author{D. Vick}
\affiliation{Quantum and Nanotechnologies Research Center, National Research Council, Edmonton, Alberta T6G 2M9, Canada}

\author{A. C\'ardenas}
\affiliation{Instituto de Ciencias de Materiales de Madrid, Consejo Superior de Investigaciones Cientificas, Madrid 28049, Spain}

\author{N. Haram}
\affiliation{Joint Attosecond Science Laboratory, National Research Council and University of Ottawa, Ottawa, Ontario K1N 5A2, Canada}

\author{P. Bastani}
\affiliation{Joint Attosecond Science Laboratory, National Research Council and University of Ottawa, Ottawa, Ontario K1N 5A2, Canada}
\affiliation{Advanced Laser Light Source, Institut National de la Recherche Scientifique, 1650 Boulevard Lionel-Boulet, Varennes, Qu\'ebec J3X 1P7, Canada}

\author{S. Gholam-Mirzaei}
\affiliation{Joint Attosecond Science Laboratory, National Research Council and University of Ottawa, Ottawa, Ontario K1N 5A2, Canada}

\author{S. Mokhtari}
\affiliation{Joint Attosecond Science Laboratory, National Research Council and University of Ottawa, Ottawa, Ontario K1N 5A2, Canada}
\affiliation{Advanced Laser Light Source, Institut National de la Recherche Scientifique, 1650 Boulevard Lionel-Boulet, Varennes, Qu\'ebec J3X 1P7, Canada}

\author{V. Jelic}
\affiliation{Joint Attosecond Science Laboratory, National Research Council and University of Ottawa, Ottawa, Ontario K1N 5A2, Canada}
\affiliation{Advanced Laser Light Source, Institut National de la Recherche Scientifique, 1650 Boulevard Lionel-Boulet, Varennes, Qu\'ebec J3X 1P7, Canada}

\author{J. Chen}
\author{J. Canlas}
\author{J. Tordiff}
\author{Md. W. Rahman}
\affiliation{Quantum and Nanotechnologies Research Center, National Research Council, Edmonton, Alberta T6G 2M9, Canada}

\author{A. Yu. Naumov}
\author{D. M. Villeneuve}
\author{A. Staudte}
\affiliation{Joint Attosecond Science Laboratory, National Research Council and University of Ottawa, Ottawa, Ontario K1N 5A2, Canada}

\author{M. Salomons}
\affiliation{Quantum and Nanotechnologies Research Center, National Research Council, Edmonton, Alberta T6G 2M9, Canada}

\author{R. E. F. Silva}
\affiliation{Instituto de Ciencias de Materiales de Madrid, Consejo Superior de Investigaciones Cientificas, Madrid 28049, Spain}
\affiliation{Max-Born-Institut for Nonlinear Optics and Short Pulse Spectroscopy, Berlin D-12489, Germany}
\author{\'A. Jim\'enez-Gal\'an}
\affiliation{Instituto de Ciencias de Materiales de Madrid, Consejo Superior de Investigaciones Cientificas, Madrid 28049, Spain}
\affiliation{Max-Born-Institut for Nonlinear Optics and Short Pulse Spectroscopy, Berlin D-12489, Germany}

\author{G. Vampa}
\affiliation{Joint Attosecond Science Laboratory, National Research Council and University of Ottawa, Ottawa, Ontario K1N 5A2, Canada}

\begin{abstract}
	High-harmonic generation (HHG) and tunnel ionization in solids are strong-field phenomena sensitive to the attosecond dynamics of laser-driven electron-hole pairs. While HHG has been used extensively as a spectroscopic probe in crystalline systems, its potential as a probe of disorder on the length scale of electron-hole trajectories remains relatively unexplored. Furthermore, with the advent of petahertz electronics, there is a growing need for controlling ionization on the nanoscale. Here, we show how focused-ion beam driven amorphization of silicon leads to a giant enhancement ($>$250 times) of yield for lower-order harmonics and quenching of higher-order harmonics, consistent with a dramatically enhanced tunnel ionization yield and disorder-induced decoherence, respectively. Modelling the real-space quantum dynamics, we show how electron-hole coherence in the amorphous phase is lost over a characteristic length of approximately five lattice sites. HHG spectroscopy also reveals remnant order inherited from the original cubic structure. Our results establish HHG spectroscopy as a probe of medium-range order and pave the way for new opportunities in lightwave nanoelectronics.
\end{abstract}

\maketitle

Solid-state strong-field physics is a growing discipline with applications ranging from attosecond spectroscopy \cite{heide_ultrafast_2024} to lightwave electronics \cite{borsch_lightwave_2023,heide_petahertz_2024}. In semiconductors, radiation from high-harmonic generation (HHG) can be used to monitor ionization, coherent transport, and recollision of electron-hole pairs as they explore the crystal energy landscape \cite{vampa_linking_2015,schubert_sub-cycle_2014,yue_introduction_2022,takeda_ultrafast_2024}. The merger of strong-field physics and photonics also provides a plethora of tools to control HHG and ionization dynamics \cite{ciappina_attosecond_2017,korobenko_insitu_2022,li_highorder_2023,purschke_microscopic_2023,roscam_abbing_enhancing_2024}. Nevertheless, only a few studies have focused on how chemical or structural modifications can be used to control strong-field processes\cite{sivis_tailored_2017,huang_high-order-harmonic_2017,yu_enhanced_2019,nefedova_enhanced_2021}. Furthermore, while liquid-phase HHG has recently emerged as a powerful probe of scattering \cite{alexander_observation_2023,mondal_high-harmonic_2023,mondal_multi-plateau_2026}, solid-HHG spectroscopy has focused primarily on crystalline (ordered) systems with few exceptions \cite{you_high-harmonic_2017,annunziata_high-order_2024,korolev_tunable_2024}. 

Amorphous solids are a close analogy to liquids, however, they can display a variety of structural motifs. In fact, the true nature of the underlying structure in amorphous silicon, one of the most well studied and technologically relevant disordered solids, has been debated for decades. Various measurements have pointed to either a continuous random network model or paracrystalline solid \cite{Voyles2003,Treacy2012,Roorda2012PRL,Lewis2022,Rosset2025}. Indeed, more generally in materials science a significant effort has focused on probing short (1-5 \AA) to medium (5-30 \AA) range order in solids \cite{elliott_mediumrange_1991, sorensen_revealing_2020, lan_mediumrange_2021}. The trajectories of laser-driven electron-hole pairs explore the energy landscape across several nanometers, suggesting that recolliding electron-hole pairs should be sensitive to medium-range order. Furthermore, amorphous silicon displays an unusually large difference in band gap relative to its crystalline phase \cite{tauc_optical_1968,kramer_electronic_1971,pierce_electronic_1972}, suggesting that amorphization should dramatically modify ionization. 

Here, we show how amorphization of Si driven by a high-dose gallium focused-ion beam ($\mathrm{Ga^+}$ FIB) irradiation reshapes the HHG spectrum, dramatically enhancing the efficiency of lower-order harmonics while suppressing the emission of higher-order harmonics. These results are consistent with a giant enhancement of tunnel ionization yield combined with a disorder-driven decoherence that suppresses recollision from trajectories with large excursions. Furthermore, HHG reveals hidden structure remnant from the original cubic crystal, suggesting a statistical preference for medium-range order to remain aligned with the original lattice. Our conclusions are supported by a minimal model of the real-space quantum dynamics that captures the effect of disorder on the coherence of electron-hole pairs. This approach, which holds in the pertinant mixed-regime of light-matter interaction, reproduces experimentally-measured spectra and allows us to determine the characteristic length scale over which coherence is lost due to the amorphous structure. Our work paves the way for HHG spectroscopy as a quantitative probe of medium-range order in amorphous solids and enables control over lightwave-driven charge injection in Si with nanoscale precision.

\begin{figure}
	\includegraphics{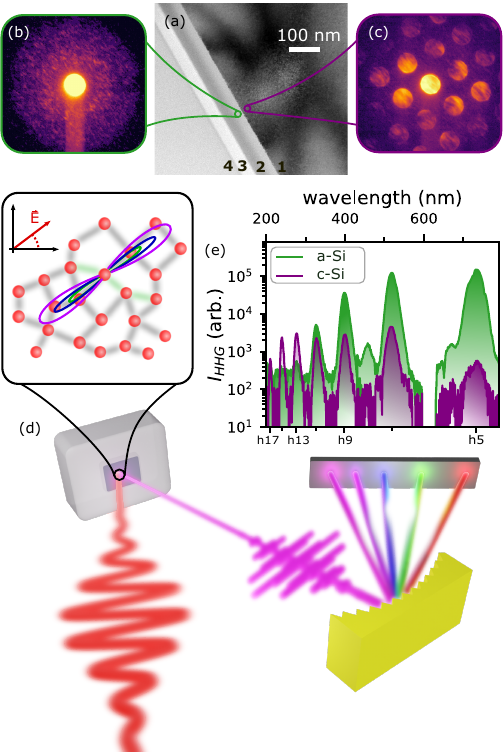}
	\caption{\label{fig:intro}HHG in amorphized silicon. (a) TEM cross section of a lamella extracted from a c-Si sample treated with Ga$^+$ FIB with an ion dose of $\mathrm{5\times 10^{15}\,cm^{-2}}$.	CBED patterns acquired (b) in the amorphous region near the surface and (c) immediately below the amorphous layer. The colormaps are logarithmically scaled to enhance contrast. (d) Schematic layout of the experiment. A strong laser field drives HHG in the sample, which is collected in reflection mode and directed to a UV-VIS spectrometer. The inset depicts the semiclassical trajectories of 3 different harmonic orders with increasing energy from green-blue-violet. (e) HHG spectra measured from c-Si (purple) and a-Si (green) with a peak laser intensity of $\mathrm{360 \, GW\cdot cm^{-2}}$.}
\end{figure}

A cross section of the amorphized Si (a-Si) surface, measured by transmission electron microscope (TEM), is shown in Fig. \ref{fig:intro}(a). The surface has four distinct layers: (1) the crystalline Si (c-Si) substrate, (2) the amorphized surface, and (3-4), protective layers deposited for lamella milling. Convergent beam electron diffraction (CBED) measurements were used to isolate the structure of the a-Si film and c-Si substrate, as shown in Figs. \ref{fig:intro}(b-c), respectively. The diffraction pattern in the a-Si layer shows two diffuse rings, as expected for a-Si \cite{edmondson_amorphization_2009}, verifying that long-range order is quenched by high-dose ion implantation. Conversely, CBED acquired beyond the a-Si boundary displays many peaks, as shown in Fig. \ref{fig:intro}(c), characteristic of a crystalline structure. As shown in Fig. \ref{fig:intro}(d), the sample is excited by mid-infrared ultrashort pulses with a 3.6 \um{} central wavelength and HHG radiation is collected in reflection mode. The laser and HHG apparatus are discussed in detail in refs. \cite{purschke_inlinedelay_2024,purschke_hhg_2026}. As illustrated schematically in the inset, the strong laser field launches electron and hole pairs along controlled trajectories. Within the class of short-trajectories relevant to solid-state HHG, electron-hole pairs for higher-order harmonics experience larger excursions and explore more of the energy landscape \cite{vampa_semiclassical_2015}.

HHG spectra are shown for a-Si (green) and c-Si (purple) in Fig. \ref{fig:intro}(e). An intensity near the damage threshold of the a-Si sample was chosen to ensure measurements were performed in the non-perturbative regime. Clearly, amorphization leads to a dramatic reshaping of the HHG spectrum. The yield of lower-order harmonics (up to h11) are \textit{enhanced} relative to c-Si and, in the case of h5, the signal is remarkably more than 250$\times$ that of c-Si. This is in stark contrast to previous studies of chemically and structurally modified semiconductors, where only minor changes in HHG were observed \cite{sivis_tailored_2017,nefedova_enhanced_2021,journigan_high_2024,korolev_tunable_2024}. Simultaneously, the yields of h13 and higher are quenched and, furthermore, in this sample h15 and h17 were not visible above the noise level despite being present at this intensity in c-Si. We note that annealing restores the c-Si like HHG spectra, suggesting that structure as opposed to chemistry is the dominant mechanism at play here \cite{purschke_hhg_2026}.

\begin{figure}
	\includegraphics{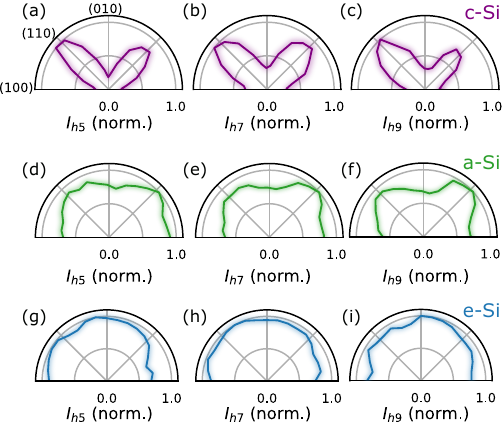}
	\caption{\label{fig:pol_dep}HHG reveals hidden cubic order. Yields of (a-c) HHG from c-Si, (d-f) a-Si, and (g-i) e-Si for the 5$^{th}$, 7$^{th}$, and 9$^{th}$ harmonics, respectively. Columns are a fixed harmonic order while rows are individual samples. The peak intensity was $360 \,\mathrm{GW\cdot cm^{-2}}$ in each measurement.}
\end{figure}

The nature of the strong-field process, which launches electron-hole pairs along prescribed directions aligned with the laser field, also enables us to probe the underlying symmetry of the structure \cite{you_anisotropic_2017,langer_symmetry-controlled_2017}. As shown in Fig. \ref{fig:pol_dep}(a-c) for h5-h9, c-Si displays sharp peaks in the HHG yield from electron-hole pairs driven along the (110) direction. In an amorphous phase consistent with a continuous random network, an isotropic HHG yield with polarization is expected. As shown in Fig. \ref{fig:pol_dep}(d-f) for h5-h9, the sharp peaks in the polarization-dependent yield are indeed lost, however, the measurement still contains some signature of hidden structural correlation that follows the original cubic lattice (recall that the CBED pattern in Fig. \ref{fig:intro}(b) does not show residual cubic structure). As shown in the Supplemental Material, this remnant structure cannot be due to contributions from the buried c-Si substrate, therefore, the structure must arise within the a-Si film itself \cite{purschke_supp}.

In contrast, as seen in Fig. \ref{fig:pol_dep}(g-i), no such evidence of cubic structure exists in HHG from an amorphous Si film prepared by thermal evaporation (e-Si) as opposed to ion implantation. Clearly, this difference originates from the difference in  sample preparation. The original crystalline structure provides a template that leads to a statistical preference for alignment of short- or medium-range order that follows the original c-Si lattice. Polarization-dependent HHG is sensitive to this preferential alignment due to  anisotropic ionization or trajectory effects. A closer look at Figs. \ref{fig:pol_dep}(d-f) reveals that the peaks become sharper with increasing harmonic order. As we show later, this suggests that trajectory effects arising from medium-range order are the likely origin of the observed structure.

\begin{figure}
	\includegraphics{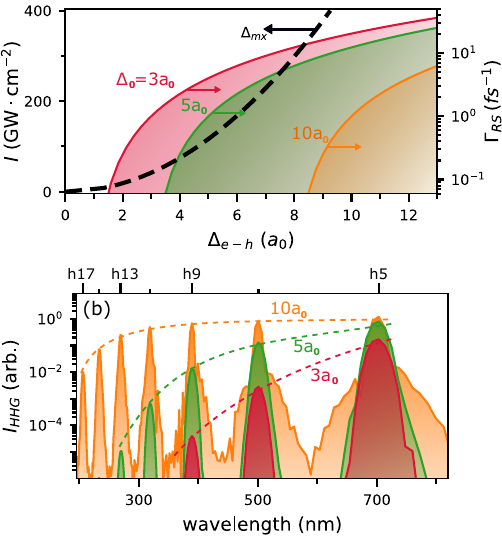}
	\caption{\label{fig:real_space_deco}Quantum dynamics of disorder-driven decoherence. (a) Real-space dephasing rate, $\Gamma_{RS}$, as a function of site-to-site separation for 3 different parameterizations. The curves are indexed by $\Delta_0$, as defined in the text. The black curve is a semiclassical calculation of the peak electric field required for recolliding electron-hole pairs to reach a maximum separation of $\Delta_{e-h}$. (b) Calculated HHG spectra for a-Si (band gap 1.75 eV) for the three different parameterizations of $\Gamma_{RS}$ outlined in (a). All calculations used a peak laser intensity of 340 $\mathrm{GW\cdot cm^{-2}}$.}
\end{figure}

Here, we shift focus to modelling our observations. These key observations in Fig. \ref{fig:intro}(e) - namely, the giant enhancement of lower-order harmonic yield and quenching of higher-order harmonics - arise due to competing effects associated with amorphization of the structure. First, introducing disorder relaxes momentum conservation laws such that a-Si has a direct band gap of 1.75 eV, effectively cutting the tunneling barrier in half \cite{tauc_optical_1968}. Second, the destruction of long-range order introduces irregularity on nanometer length scales. We estimated the maximum excursion length, $\Delta_{mx}$, of the highest-energy recolliding trajectories as a function of intensity, as shown in Fig. \ref{fig:real_space_deco}(a), black line \cite{vampa_semiclassical_2015}. Indeed, electron-hole pairs explore the energy landscape over several lattice sites, thus, disorder can randomize their return time and energy and, potentially, prevent recollision altogether. 

This trajectory picture provides a qualitative understanding \cite{you_high-harmonic_2017}, and has even been used quantitatively to extract the scattering length from liquid-phase HHG \cite{mondal_high-harmonic_2023}. However, as is often the case for solid-state attosecond science, our experiments operate in the mixed multiphoton/tunnel ionization regime with a Keldysh parameter of $\sim 1.3$, where non-adiabatic tunneling governs ionization. As a result, fully quantum simulations are needed to accurately capture both the ionization dynamics and trajectory effects. 

The real-space picture of the semiconductor Wannier equations (SWEs, see End Matter and ref. \citenum{molinero_semiconductor_2026}) is key to our approach, as site-to-site coherences provide a quantum dynamical analog to trajectories \cite{brown_real-space_2024}. In a-Si, the short-range order (bond angle and length) resembles that of c-Si, thus, as a starting point, we consider a minimal model consisting of a periodic 1D chain of atoms. The SWEs facilitate introducing a real-space dephasing, $\Gamma_{RS}$, which was recently used to mimic the macroscopic effect of transverse phase-matching as an alternative to microscopic ultrafast dephasing \cite{brown_real-space_2024}, for which the underlying mechanism has been debated extensively \cite{floss_ab_2018,abadie_spatiotemporal_2018,orlando_simple_2020,heide_probing_coherence_2022,nagai_effect_2023,cardenas_effects_2026,mokhtari_phonon-driven_2026}. Here, we show how real-space dephasing can be used to capture \textit{real microscopic physics}: disorder-driven decoherence.

\begin{figure}
	\includegraphics{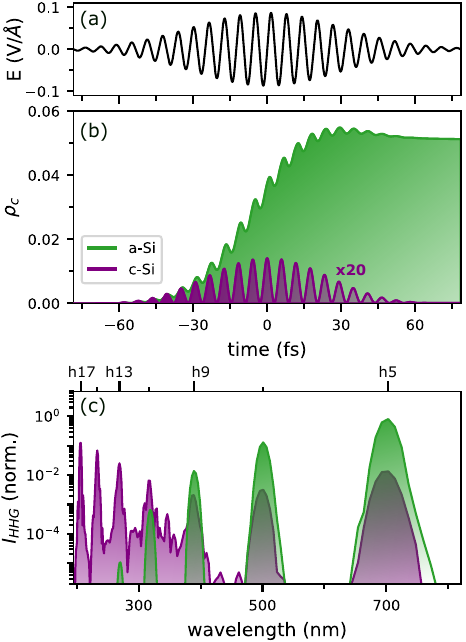}
	\caption{\label{fig:calcs} Giant enhancement of tunnel-ionization yield. (a) Applied electric field waveform and (b) conduction band population as a function of time for a-Si (green) and c-Si (purple). Note that the c-Si population is scaled by a factor of 20. Calculations are performed using the model described in the text with a band gap of 3.5 eV for c-Si and 1.75 eV for a-Si. The peak intensity was 340 $\mathrm{GW\cdot cm^{-2}}$ for each calculation. (c) The corresponding HHG spectra. The a-Si (green) curve was calculated with characteristic length $\Delta_0=5a_0$. }
\end{figure}

Focusing first on the ionization yield, we simulated the strong-field quantum dynamics using the SWEs with a multi-cycle driving field, as shown in Fig. \ref{fig:calcs}(a) \cite{molinero_semiconductor_2026}. The conduction-band populations as a function of time are shown in Fig. \ref{fig:calcs}(b) using models parameterized to yield direct band gaps equal to that of c-Si (3.5 eV, purple) and a-Si (1.75 eV, green). The final density in the a-Si model calculation is 2 orders of magnitude larger than in c-Si. Furthermore, the population in the c-Si case is primarily virtual (note the deep oscillations) \cite{mcdonald_intense-laser_2017,Boroumand2025StrongField} while in the a-Si case, the population shows more step-like increases reminiscent of tunnel ionization \cite{schultze_attosecond_2014}. The ionization yields were relatively insensitive to $\Gamma_{RS}$ (see Supplemental Material \cite{purschke_supp}). The corresponding HHG spectra are shown in Fig. \ref{fig:calcs}(c) and, indeed, the low-order harmonic yield is comparably higher in a-Si vs c-Si to what is observed in the experiment. Thus, the decreased gap of a-Si dramatically enhances ionization and low-order harmonic generation. With FIB, amorphous regions can be patterned to $\sim$20 nm length scales \cite{drezner_high_2017}, enabling the creation of ultrasmall active regions for charge injection by tunnel ionization.

Next, we explore the effect of real-space dephasing on the HHG spectra. Fig. \ref{fig:real_space_deco}(b), shows HHG spectra in a-Si calculated using three different parameterizations of $\Gamma_{RS}$, corresponding to the three curves in Fig. \ref{fig:real_space_deco}(a). Clearly, the roll-off in the HHG spectra is sensitive to the location of the real-space dephasing boundary, $\Delta_0$, defined as the location where the dephasing rate reaches 3 times the drive laser frequency (see End Matter). This definition was chosen so that calculated spectra were sensitive primarily to the characteristic length, $\Delta_0$, as opposed to the functional form of $\Gamma_{RS}$ (see Supplemental Material \cite{purschke_supp}). 

As decoherence on such short length scales is intrinsically linked to disorder, we conjecture that $\Delta_0$ acts as a proxy for medium-range order, which is then imprinted onto the roll-off in the HHG spectra. This evokes a coherent-transport analogy to time-resolved terahertz spectroscopy, where carrier localization in the diffusive regime manifests in the conductivity spectra \cite{cocker_microscopic_2017,purschke_ultrafast_2021}. Closer inspection of Fig. \ref{fig:real_space_deco}(b) in comparison to Fig. \ref{fig:intro}(e), we find that $\Delta_0\approx 5a_0$ is required to reproduce experimental spectra, where $a_0=5.43\,\text{\AA}$ is the silicon lattice constant. Interestingly, real-space dephasing also modifies the power-law scaling of each harmonic, providing a useful corroboration of the conclusion drawn from the spectra (see Supplemental Material \cite{purschke_supp}). 

We briefly return to the remnant structural correlation observed in Fig. \ref{fig:pol_dep}(d-f). As a starting point, one could consider the a-Si sample as an ensemble of 1D chains oriented at random angles. Clearly, this approach could not reproduce the observed polarization dependence of HHG. Moving a step further, one could consider an ensemble of orientations with a slight preference to lie along the original (110) direction, however, this would not capture the sharpening of polarization dependence with increasing harmonic order. Instead, our observations are consistent with an isotropic distribution of chains with an angle-dependent characteristic length, $\Delta_0$, and thus dephasing rate, implying a slight statistical preference for medium-range order aligned with the original cubic lattice. 

To conclude, our results provide a blueprint for future studies using HHG as a spectroscopic probe of nanoscale disorder that can provide unique information about the structure. We anticipate our work to spur new insight into the nature of attosecond decoherence arising from both static and dynamical disorder, such as phonons \cite{mokhtari_phonon-driven_2026} and zero-point motion \cite{cardenas_effects_2026} or in liquid-phase HHG in the non-adiabatic regime \cite{alexander_observation_2023,mondal_high-harmonic_2023}. Furthermore, we demonstrate that amorphization leads to a giant enhancment in tunnel ionization yield, increasing the HHG yield orders of magnitude more than previously observed for chemical or structural modification of solids. In silicon, structural disorder acts as an on-off switch that can be patterned on ultrasmall length scales using FIB or other CMOS compatible technologies. This will enable the fabrication of nanoscale circuit elements for devices in lightwave electronics, where carriers can be collected faster than they are trapped.

\section{End Matter}
\subsection{Real-space quantum dynamics}
To capture the essential physics, we employ a minimal tight-binding one-dimensional chain with two orbitals per unit cell separated by a distance $a_0/2$, where $a_0$ is the unit cell vector of c-Si ($a_0=5.43$ $\text{\AA}$). Both orbitals have different on-site energies, resulting in a band gap of $E_{g,\text{a-Si}}=1.75$eV and $E_{\text{g,c-Si}}=3.5$ eV for the a-Si and c-Si, respectively \cite{tauc_optical_1968}.

The HHG spectra is obtained using the methodology outlined in Molinero et al. \cite{molinero_semiconductor_2026}. In particular, we solve the equation of motion of the electronic density matrix in real space, within the dipole approximation and in the length gauge. The effects of a shorter coherence length in the HHG spectra is modeled by using a distance-dependent decoherence term that suppresses quadratically coherences between sites. The dephasing rate has the simple functional form,
\begin{equation}
\Gamma_{rs}(x)=\begin{cases}
0 & x<\delta_{0}\\
\beta\left(x-\delta_{0}\right)^{2} & x\geq\delta_{0}
\end{cases}
\end{equation}
In the text, we fix the curvature $\beta$ by requiring that $\beta = 3\omega_0/a_0$ and we change the parameter $\Delta_0 = a_0+\delta_0$. In the Supplemental Material we show that the harmonic roll-off is independent of $\beta$ \cite{purschke_supp}.

\bibliographystyle{apsrev4-2}
\bibliography{Giant_enhancement.bib}

\end{document}